\documentclass[preprint,12pt]{elsarticle}

\usepackage{lineno,hyperref}
\usepackage{amsmath}
\usepackage{graphicx}
\usepackage[labelsep=period,labelfont=bf,figurename=figure]{caption}
\usepackage{subcaption}
\usepackage{float}
\journal{TBD}
\begin{document}
	
	\begin{frontmatter}
		
		\title{Microscopic Black Hole Events in Future Hadron Colliders}
		
		\author{Halil Gamsızkan\corref{mycorrespondingauthor}}
		\address{Eskişehir Technical University, Eskişehir, Turkey}
		
		\begin{abstract}
Microscopic black hole production at future hadron colliders is a promising avenue to explore low scale gravity models like ADD. This study investigates production cross-sections and final state properties of these black holes using the BlackMax event generator. We analyze scenarios with varying proton-proton collision energies (27 TeV and 100 TeV), number of extra dimensions (2, 4, and 6), and black hole rotation. Black holes are found to decay to up to seventeen particles, which have high transverse momentum (up to 30 TeV) and concentrated in the central region. This unique signature allows them to be distinguished from background processes. The parameter space explored suggests future colliders like HE-LHC and FCC-hh have the potential to probe low scale gravity models.
\end{abstract}		
		
	\end{frontmatter}
	
	
	\section{Introduction}
		At the Large Hadron Collider (LHC), searches for signals of large extra dimensions and microscopic black hole (BH) production have so far resulted with null results. Limits on the fundamental Planck scale $M_{\rm D}$ and minimum BH mass $M_{\rm th}$ have been set at values close to 10 TeV \cite{RAPPOCCIO2019100027, cmsBHSph2018}. If the fundamental Planck scale is just above these energies, it may be possible to discover extra dimensions through BH production at future colliders. 

        Though the discovery of the Higgs boson in the LHC experiments stands as a crowning achievement, many important questions in fundamental physics remain to be answered. Answering these questions may require more energetic and luminous colliders, so there are plans to  construct future hadron colliders. Among these are the High Energy LHC (HE-LHC) and the Future Circular Collider (FCC-hh) projects. HE-LHC is designed to deliver proton-proton collisions at $\sqrt{s}=27$ TeV energies and deliver a total integrated luminosity of 10 ${\rm ab}^{-1}$ \cite{Abada2019}. FCC-hh is the hadron collider of the FCC project expected to deliver proton-proton collisions at $\sqrt{s}=100$ TeV and a total integrated luminosity of $\sim 20\; {\rm ab}^{-1}$ \cite{Abada2019b}. 
        
        The hierarchy problem, which may be stated as the large hierarchy between the EW scale ($\sim100$ GeV) and the Planck scale ($\sim10^{18}$ GeV), motivated a number of models with extra dimensions as solutions. One such model is the Arkani-Hamed–Dimopoulos–Dvali (ADD) large extra dimension model \cite{ANTONIADIS1998257, ARKANIHAMED1998263, PhysRevD.59.086004}. In this model, the existence of $n$ large compact extra dimensions is proposed. Standard Model (SM) particles are confined to the three dimensional ``brane" while only gravitons are allowed to propagate to the 3+$n$ dimensional ``bulk". This effectively dilutes the gravitational interaction in long distances and hence the apparent hierarchy between the Planck and electroweak scales. 
        
        Among the major parameters of the ADD model are the ``true" Planck scale $M_{\rm D}$, the number of extra dimensions $n$ and the radius of extra dimensions $r$. 
         $M_{\rm D}$ is related to 3-dimensional Planck mass $M_{\rm pl}$ as \cite{ANTONIADIS1998257}:
        \begin{equation}
            M_{\rm pl}^2=8\pi M_{\rm D}^{n+2}r^n.
        \end{equation}
        Setting $M_{\rm D}$ to values in the TeV range gives extra dimension radii values in orders of fm to mm, which are large compared to characteristic distance scales of particle interactions, hence the name {\em large} extra dimensions. We also note that the case with $n=1$ leads to $r\sim10^{13}$ cm, which would imply deviations from Newtonian gravity at the solar system scales and since no such deviations are observed this case is already excluded.
        
        Stronger gravity in short distances, as implied by the ADD model, means a BH may be formed at the collision of two particles if the hoop conjecture \cite{thorne1972,banks1999model} conditions are satisfied \cite{PhysRevLett.87.161602, PhysRevD.65.056010}. Hoop conjecture states that it's possible to form a BH with mass equal to the available center of mass energy when the impact parameter of the colliding particles is no larger than twice the extra dimensional Schwarzschild radius ($b<2r_S$) defined as \cite{cmsBHSph2018}:
        \begin{equation}
            r_S=\frac{1}{\sqrt\pi M_{\rm D}}\left[\frac{M_{\rm BH}}{M_{\rm D}} 
                \left(\frac{8\Gamma(\frac{n+3}{2})}{n+2}\right)\right]^\frac{1}{n+1},
        \end{equation}where $\Gamma$ is the gamma function. In the simplest form, the production cross-section can be written as $\sigma\approx\pi r_S^2$ \cite{PhysRevLett.87.161602}. 
        Since details of the underlying quantum gravity theory are unknown, we cannot predict the mass threshold $M_{\rm th}$ at which BH production will start, more specifically on whether this energy is the Planck scale or some higher energy. Finally, the produced BHs may be electrically charged and may have nonzero angular momentum. 
       
        Once formed, a microscopic BH is expected to evaporate promptly\footnote{A typical lifetime of a microscopic BH with mass relevant to the collision energies studied here are $\sim10^{-27}-10^{-26}$ s \cite{landsberg2015}}. BH evaporation is a multistage process, starting with the ``balding'' phase in which the BH loses its multipole moments and quantum numbers. This is followed by the spin-down phase in which it loses it's angular momentum and becomes a Schwarzschild BH; then at the Hawking stage, it thermalizes via Hawking radiation \cite{landsberg2015}. At the Planckian (final) stage of the BH evaporation, the presently unknown quantum gravity effects become important. BHs at this stage may leave a stable remnant with mass $\sim M_{\rm D}$ or decay to only a few particles (conserving gauge charges it carries) and hence fully evaporate.

        Hawking radiation is both thermal and democratic in nature. Hawking temperature of a $4+n$-dimensional BH is \cite{landsberg2015}:
        \begin{equation}
            T_H=\frac{n+1}{4\pi r_S},
        \end{equation}
        which means that there is an inverse relationship between the mass of a BH and its temperature.
        Energies of the radiated particles are determined by the Hawking temperature of the BH, while the particle types are determined by the available degrees of freedom per particle in the SM. The democratic nature of Hawking radiation has a few implications: If there are new undiscovered particles, they will be radiated through Hawking radiation, which may help with their discovery, see e.g. \cite{PhysRevLett.124.051801}. Second, since quarks and gluons have a relatively high share of degrees of freedom in the SM, BH final states are expected to be rich in hadronic activity. 

We would like to conclude this section with a few words on the safety of microscopic BH production. The safety of possible stable BH production in a future proton collider with 100 TeV energy has been studied and for the models with more than 6 dimensions Earth’s accretion times to macroscopic sizes have been found larger than the lifetime of the Solar system \cite{sokolov2017}. In the mentioned study, the safety arguments were also supported with astrophysical arguments.
       
\section{Simulations of microscopic black hole production and decay}
In this article, we aim to study microscopic BH production and decay at future hadron colliders in the ADD model with various model parameters. We use an event generator software to calculate production cross-sections and simulate both BH formation and evaporation final states. In the final states, we study only the event generator output, particle showering or detector simulations were left for further studies.

There are a number of event generators available to simulate production and decay of microscopic BHs, e.g. \texttt{BlackMax} \cite{PhysRevD.77.076007} and \texttt{Charybdis} \cite{Frost_2009}. In this study we used the \texttt{BlackMax v2.02} event generator. The particular PDF distributions we used were \texttt{CT18NNLO} \cite{PhysRevD.103.014013} which were used through the \texttt{LHAPDF6} software \cite{Buckley2015}. 

In our studies we assumed that minimum BH mass that can be produced ($M_{\rm th}$) is equal to the Planck mass ($M_{\rm D}$). Concerning $M_{\rm D}$, we chose 10 TeV and 15 TeV (HE-LHC and FCC-hh), 25 TeV and 40 TeV (FCC-hh) as our working points. The number of extra dimensions we studied were $n$=2, 4 and 6. Concerning BH rotation and energy loss, we picked three scenarios among the ones available in \texttt{BlackMax}, tensionless non-rotating BHs (NR), rotating nonsplit BHs (R), rotating BHs with mass and momentum loss and graviton emission (RL)\cite{cmsBHSph2018}. In the scenarios with mass and momentum loss during BH formation we set the mass and momentum loss factor to 0.1. We name the scenarios we studied by combining the collision CM energy and the rotation scenario code, for example a scenario of 100 TeV p-p collisions producing rotating BHs with mass and momentum loss and graviton emission will be referred to as S100$\_$RL. Concerning the number of extra dimensions, we will be reporting results for $n=2$ unless otherwise stated.

\begin{figure}
\begin{subfigure}{.5\textwidth}
  \centering
  \includegraphics[width=1\linewidth]{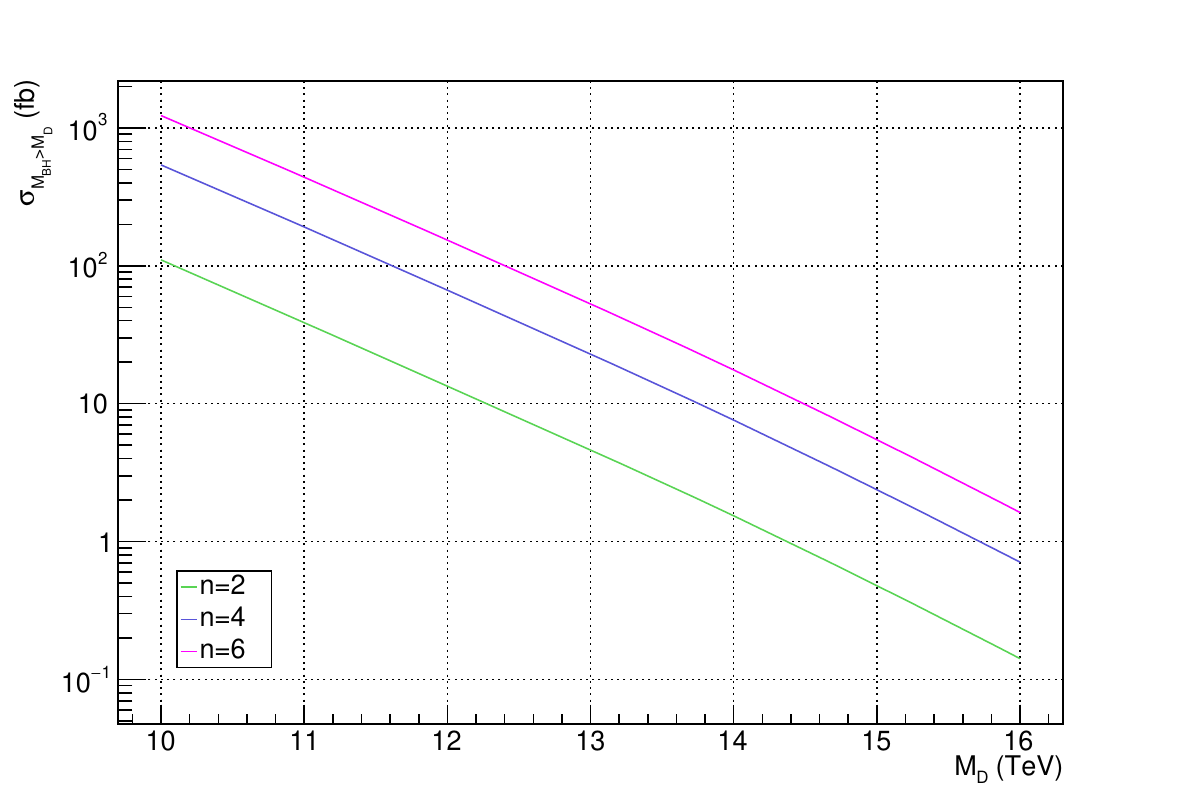}
  \caption{}
  \label{fig:xsect27TeV0}
\end{subfigure}%
\begin{subfigure}{.5\textwidth}
  \centering
  \includegraphics[width=1\linewidth]{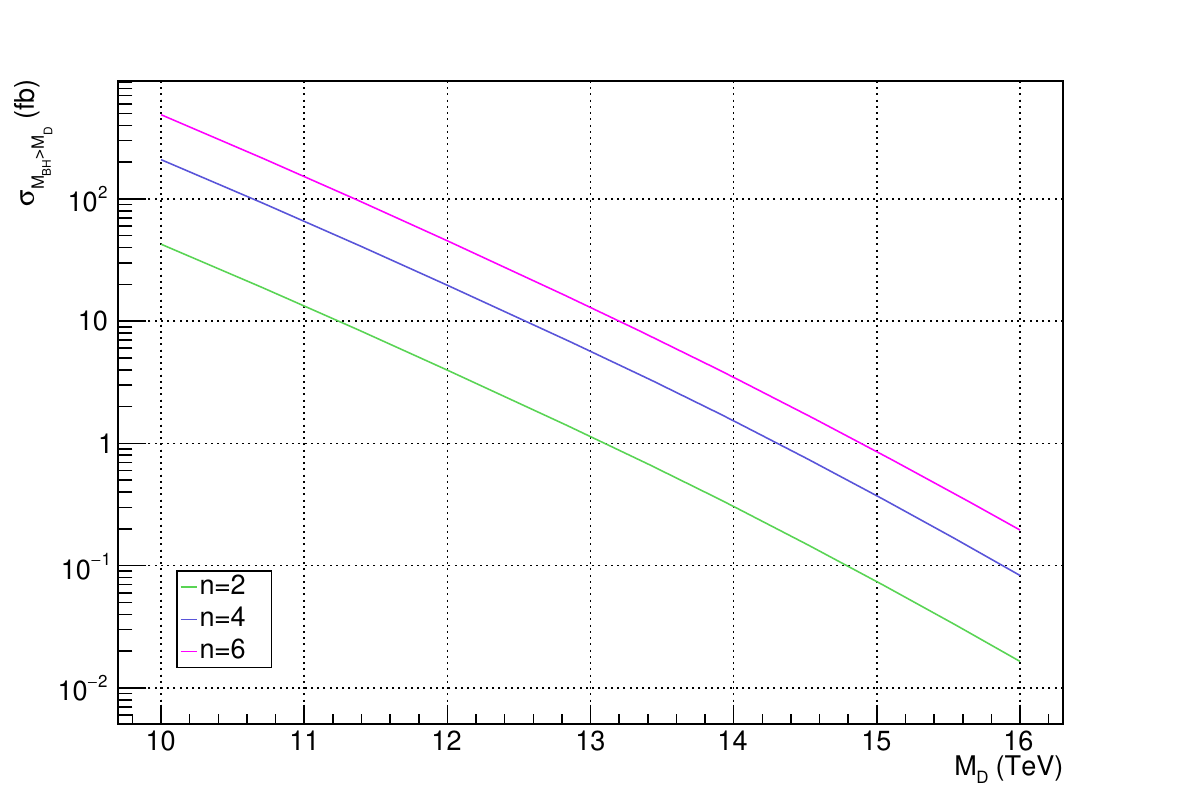}
  \caption{}
  \label{fig:xsect27TeV1}
\end{subfigure}
\caption{Cross-sections for microscopic BH production with mass higher than $M_{\rm D}$ in $\sqrt{s}=27$ TeV proton-proton collisions for (a) non-rotating BHs (scenario S27$\_$NR) and (b) for rotating BHs with mass and momentum loss factor set to 0.1 (scenario S27$\_$RL).}
\label{fig:xsect27TeV}
\end{figure}

\begin{figure}
\begin{subfigure}{.5\textwidth}
  \centering
  \includegraphics[width=1\linewidth]{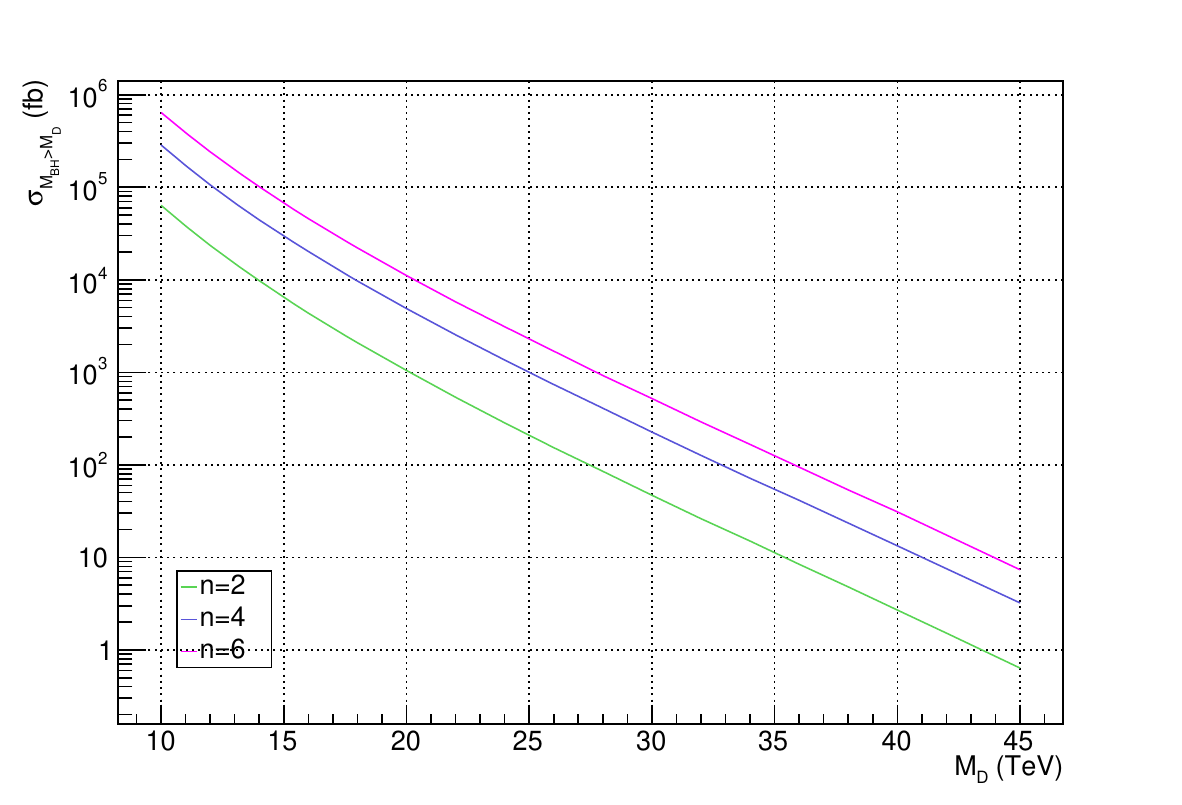}
  \caption{}
  \label{fig:xsect100TeV0}
\end{subfigure}%
\begin{subfigure}{.5\textwidth}
  \centering
  \includegraphics[width=1\linewidth]{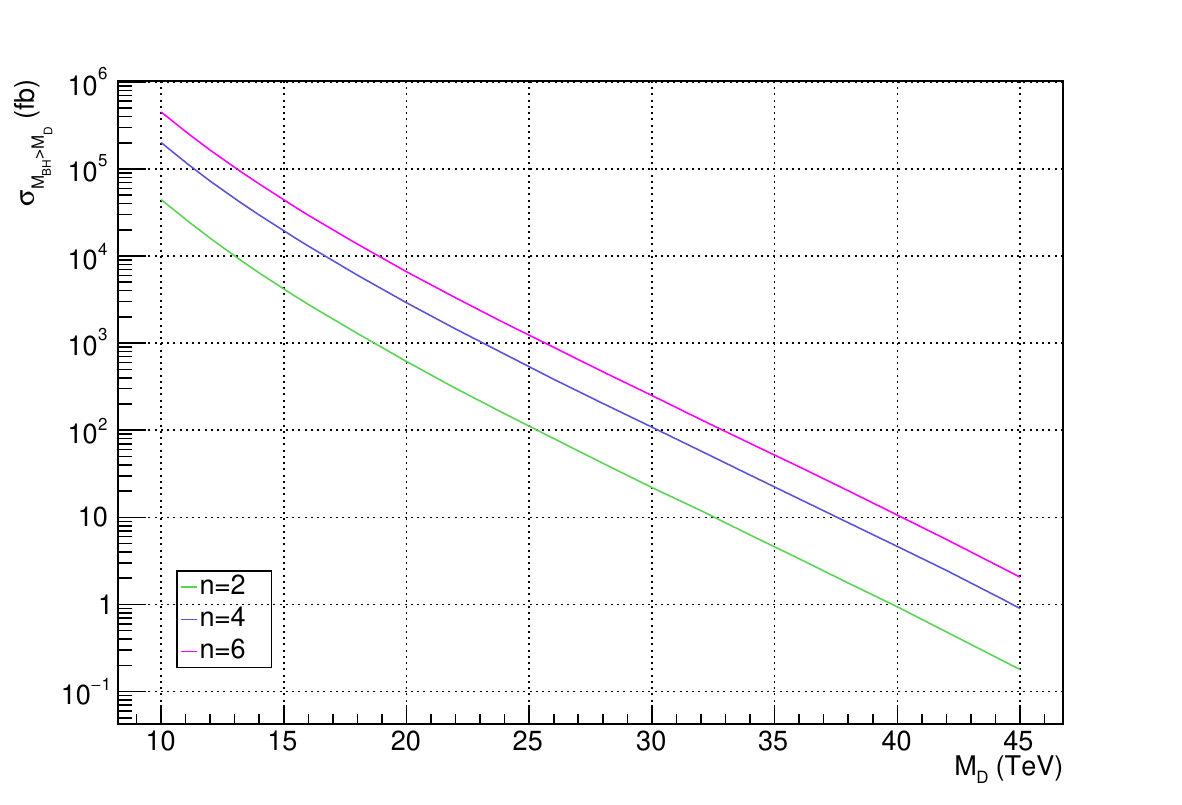}
  \caption{}
  \label{fig:xsect100TeV1}
\end{subfigure}
\caption{Cross-sections for microscopic BH production with mass higher than $M_{\rm D}$ in $\sqrt{s}=100$ TeV proton-proton collisions for (a) non-rotating BHs (scenario S100$\_$NR) and (b) for rotating BHs with mass and momentum loss factor set to 0.1 (scenario S100$\_$RL).}
\label{fig:xsect100TeV}
\end{figure}

\paragraph {Microscopic Black Hole Production}
Cross-sections for production of BHs for $27$ TeV and $100$ TeV proton collisions as calculated by \texttt{BlackMax} are shown on figure~\ref{fig:xsect27TeV} and figure~\ref{fig:xsect100TeV}. In both cases, cross-sections fall exponentially with increasing Planck mass as expected. In HE-LHC energies, cross-sections are lowered to femtobarn region at around $M_{\rm D}$ values of 15 TeV. For 100 TeV collisions this  value is around 45 TeV. We also observe that in both collision energy cases, cross-sections increase with increasing number of extra dimensions $n$ as expected. The cross-sections for producing rotating BHs (R) is the same with non-rotating BHs (NR), but the cross-sections for rotating BHs with mass and momentum loss (RL) are suppressed when compared with these cases, as there's less energy available to form the BH. We can also note that the graviton emission option does not affect the cross-sections because the BH formation related effect of this option is only to make the lost mass/momentum to be radiated in gravitons instead of photons, without affecting the formation process explicitly.

\begin{figure}
\begin{subfigure}{.5\textwidth}
  \centering
  \includegraphics[width=1\linewidth]{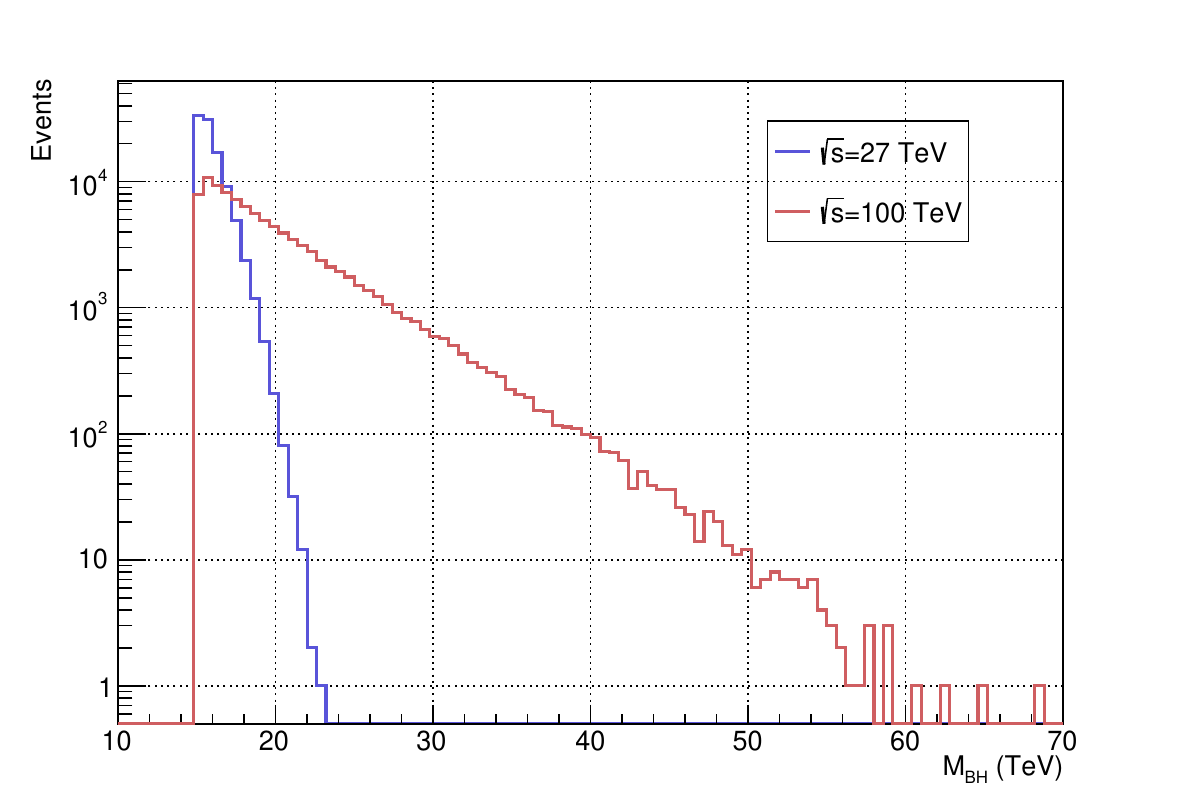}
  \caption{}
  \label{fig:mass0}
\end{subfigure}%
\begin{subfigure}{.5\textwidth}
  \centering
  \includegraphics[width=1\linewidth]{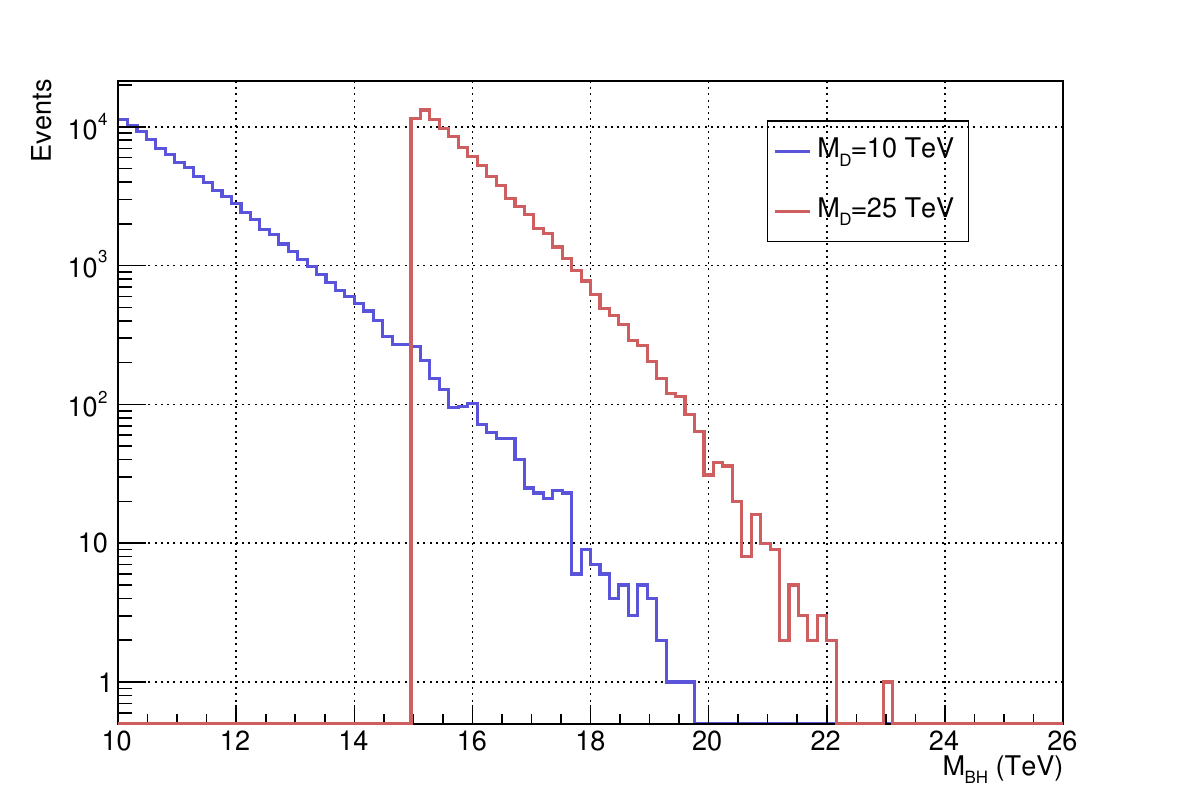}
  \caption{}
  \label{fig:mass1}
\end{subfigure}
\caption{Invariant mass distributions for (a) S27$\_$NR vs. S100$\_$NR scenarios, and (b) for S27$\_$NR scenario with  $M_{\rm D}=10$ TeV vs. $M_{\rm D}=15$ TeV. All histograms were produced with equal number of simulated events.}
\label{fig:mass}
\end{figure}

On figure~\ref{fig:mass0}, we see BH invariant mass distributions for different collision energies and for the same $M_{\rm D}$. Here, we observe that higher available energy in $100$ TeV collisions allows more massive BHs to be produced compared to the $27$ TeV case. This affects various distributions regarding the final states because the final states in $100$ TeV cases tend to come from the decay of heavier BHs.

On figure~\ref{fig:mass1}, we see mass distributions for the same collision energy but different $M_{\rm D}$. Here, we observe that with increasing $M_{\rm D}$ the mass distribution is more ``squeezed'' between $M_{\rm th}$ and the available energy. This will have important consequences in e.g. the multiplicity distributions, to be discussed below.

\begin{figure}
\begin{subfigure}{.5\textwidth}
  \centering
  \includegraphics[width=1\linewidth]{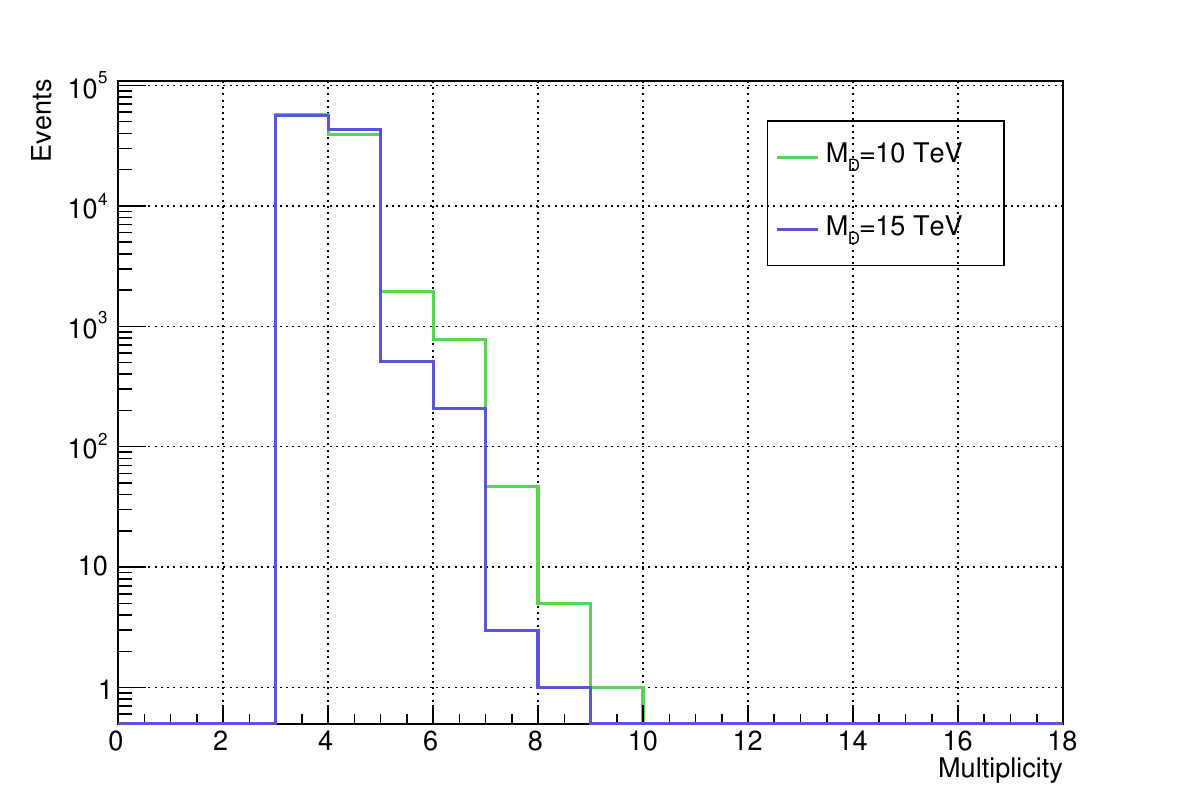}
  \caption{}
\end{subfigure}%
\begin{subfigure}{.5\textwidth}
  \centering
  \includegraphics[width=1\linewidth]{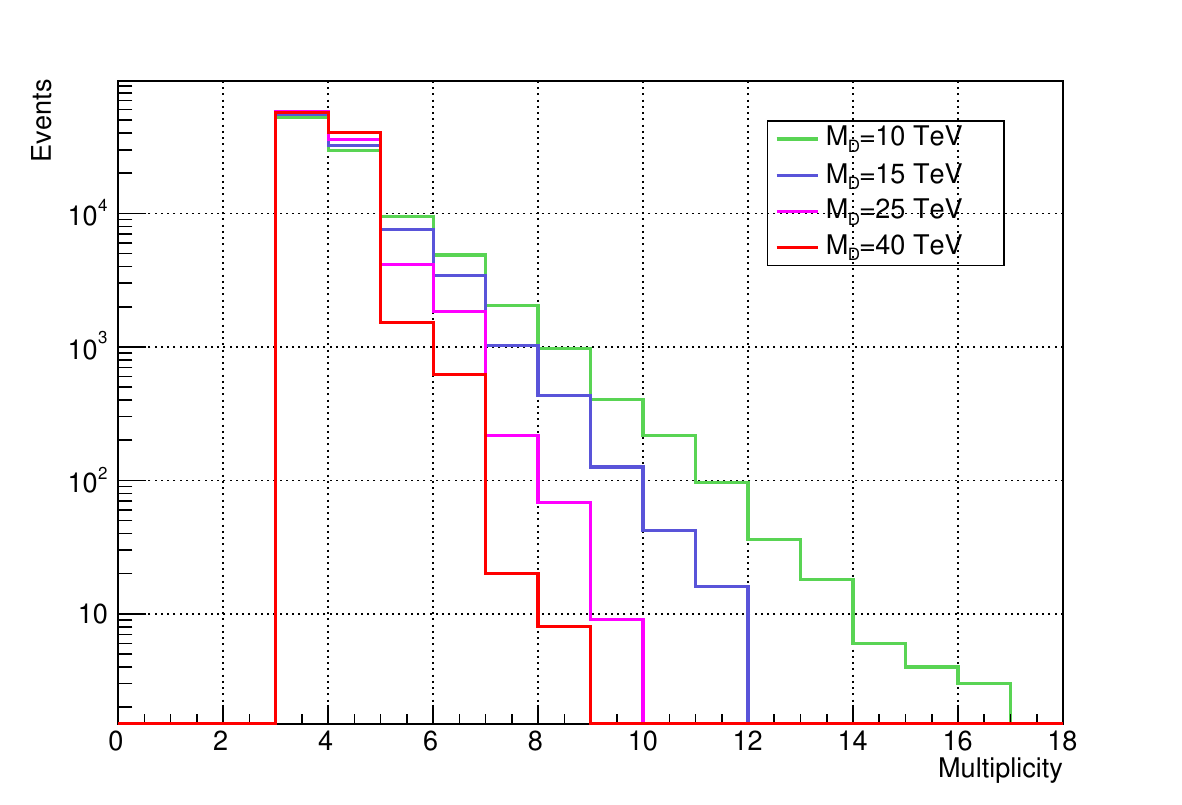}
  \caption{}
\end{subfigure}
\caption{Total number of particles emitted from non-rotating BHs during decay for (a) S27$\_$NR and (b) S100$\_$NR scenarios.}
\label{fig:multcomp}
\end{figure}

\paragraph {Microscopic Black Hole Decay}
Let's begin studying the BH evaporation final states with particle multiplicities. figure~\ref{fig:multcomp} shows distributions for the S27$\_$NR and S100$\_$NR cases. Since most BHs are produced near $M_{\rm D}$, \texttt{BlackMax} quickly switches to the Planck phase for these BHs. The scenario adopted by \texttt{BlackMax} in this stage is a final burst with minimum number of particles which conserves energy, momentum and all of the gauge quantum numbers \cite{dai2009manual}. This results in the most common multiplicity values being three and four. Heavier BHs have more energy to radiate thermally so they have higher multiplicities.
\begin{figure}
\begin{subfigure}{.5\textwidth}
  \centering
  \includegraphics[width=1\linewidth]{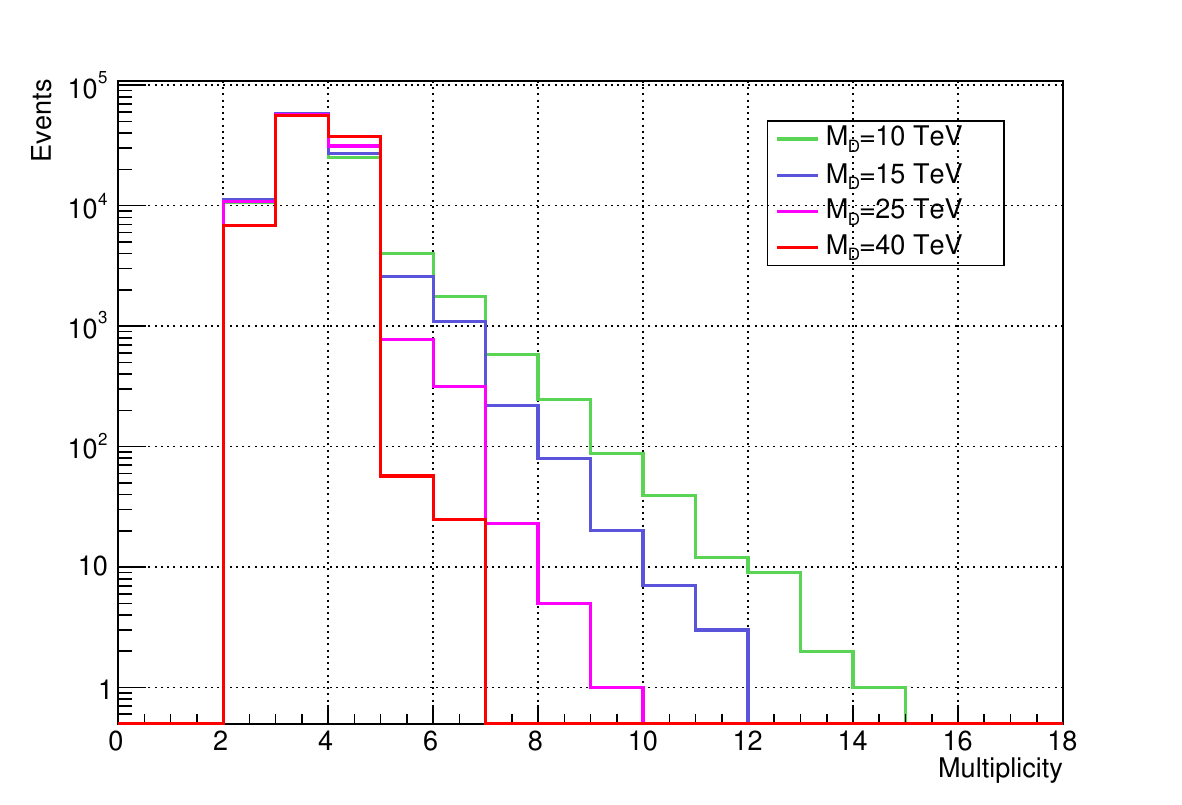}
  \caption{}
  \label{fig:comp_mult_100_r3}
\end{subfigure}%
\begin{subfigure}{.5\textwidth}
  \centering
  \includegraphics[width=1\linewidth]{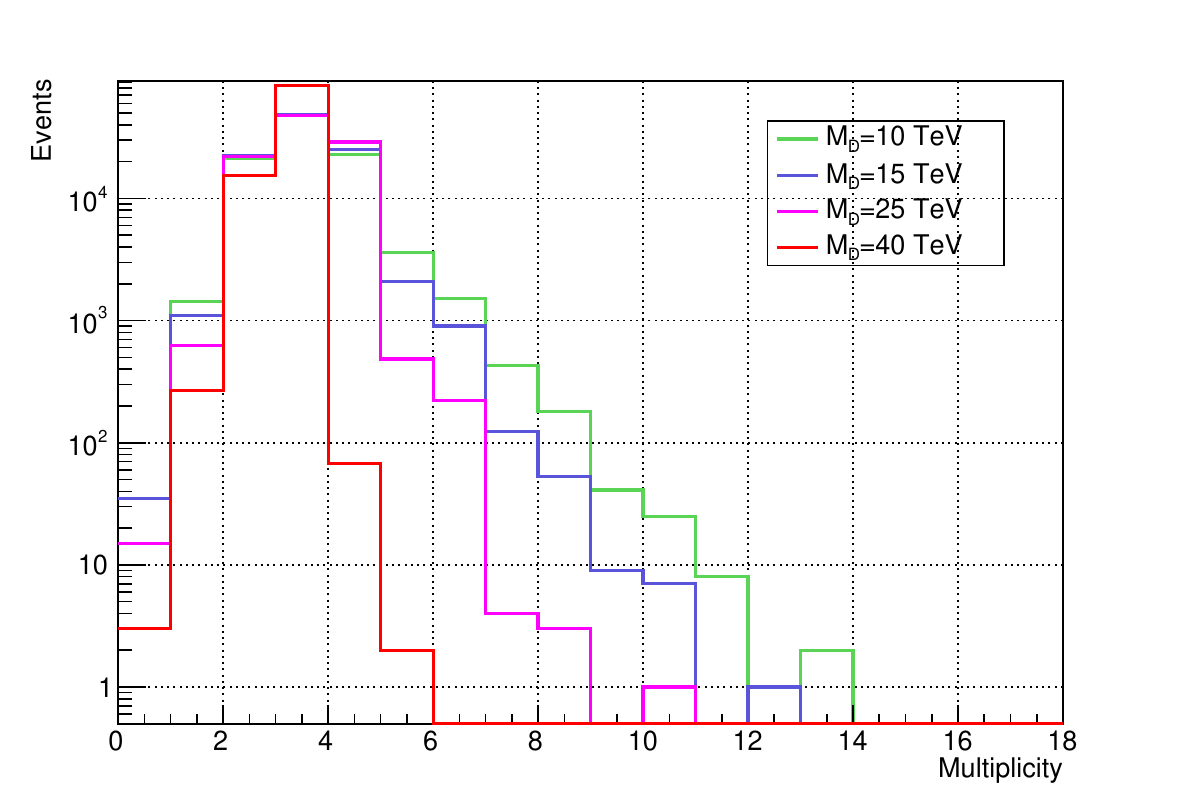}
  \caption{}
  \label{fig:comp_mult_100_r3_loss}
\end{subfigure}
\caption{Total number of particles emitted from the BH as it evaporates for (a) S100$\_$R and (b) S100$\_$RL scenarios. In the S100$\_$RL scenarios, gravitons in the final state are not counted in the multiplicities.}
\label{fig:multrcomp}
\end{figure}

In both collision energy scenarios, we observe decreasing multiplicities with increasing $M_{\rm D}$. This may initially appear counter-intuitive as heavier BHs tend to be colder, radiating softer and higher number of particles, however, as discussed before, as $M_{\rm D}$ increases the mass distribution is more ``squeezed'' between $M_{\rm D}$ and the kinematic limiting due to the collision energy (figure~\ref{fig:mass1}). As $M_{\rm D}$ increases, $M_{\rm BH}$ distribution spread becomes narrower and more BHs have masses near $M_{\rm D}$. These BHs decay with relatively low multiplicities. On the other hand, smaller $M_{\rm D}$ values mean there is more room to form BHs with masses much higher than $M_{\rm D}$. These more massive BHs will emit more particles during decay resulting in final states with higher multiplicities and this is exactly the pattern we observe here.

Concerning rotating BHs, we show results for 100 TeV collision energy scenarios on figure~\ref{fig:multrcomp}. Here at the S100$\_$R scenario (figure~\ref{fig:comp_mult_100_r3}), we observe a shift in multiplicities towards lower values compared to non-rotating scenarios. Many BHs decay to only two particles. In the S100$\_$RL scenario (figure~\ref{fig:comp_mult_100_r3_loss}), in which all the gravitons in the final state are excluded from multiplicities, we observe that some events have zero multiplicity, which means that all the energy of the BHs is emitted in the form of gravitons. These are somewhat interesting events with the energy available in the colliding partons fully disappearing in the form of gravitons with no other SM particles in the final state. A number of events with a single particle in the final state are also expected. Despite these possibilities, we see that most probable multiplicity value is three particles. At the higher end, the multiplicity distribution extends to 14 particles in the highest $M_{\rm D}$ scenario.

Number of extra dimension dependence of multiplicities have also been studied and a pattern of slightly increasing multiplicities with increasing number of extra dimensions has been observed. This increase had little effect on the expected range of multiplicities, which ranges from one particle to 17 particles among various scenarios.

\begin{figure}
\begin{subfigure}{.5\textwidth}
  \centering
  \includegraphics[width=1\linewidth]{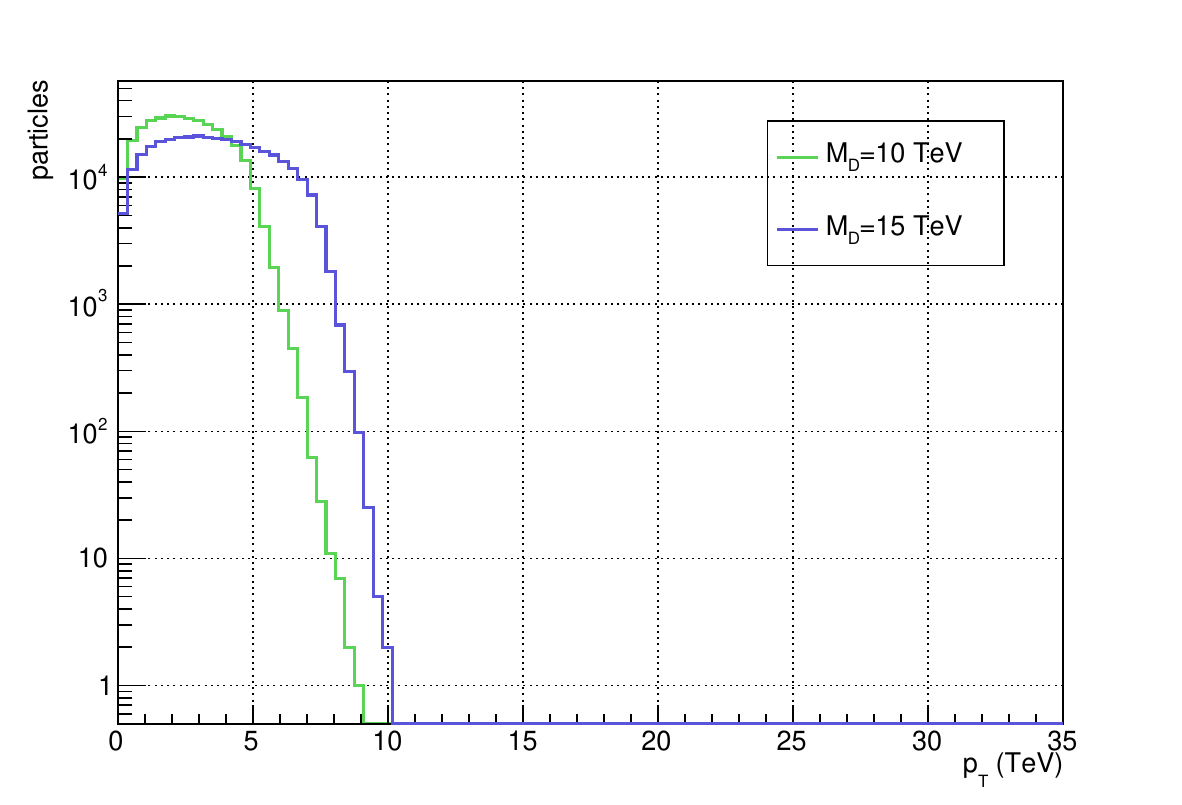}
  \caption{}
\end{subfigure}%
\begin{subfigure}{.5\textwidth}
  \centering
  \includegraphics[width=1\linewidth]{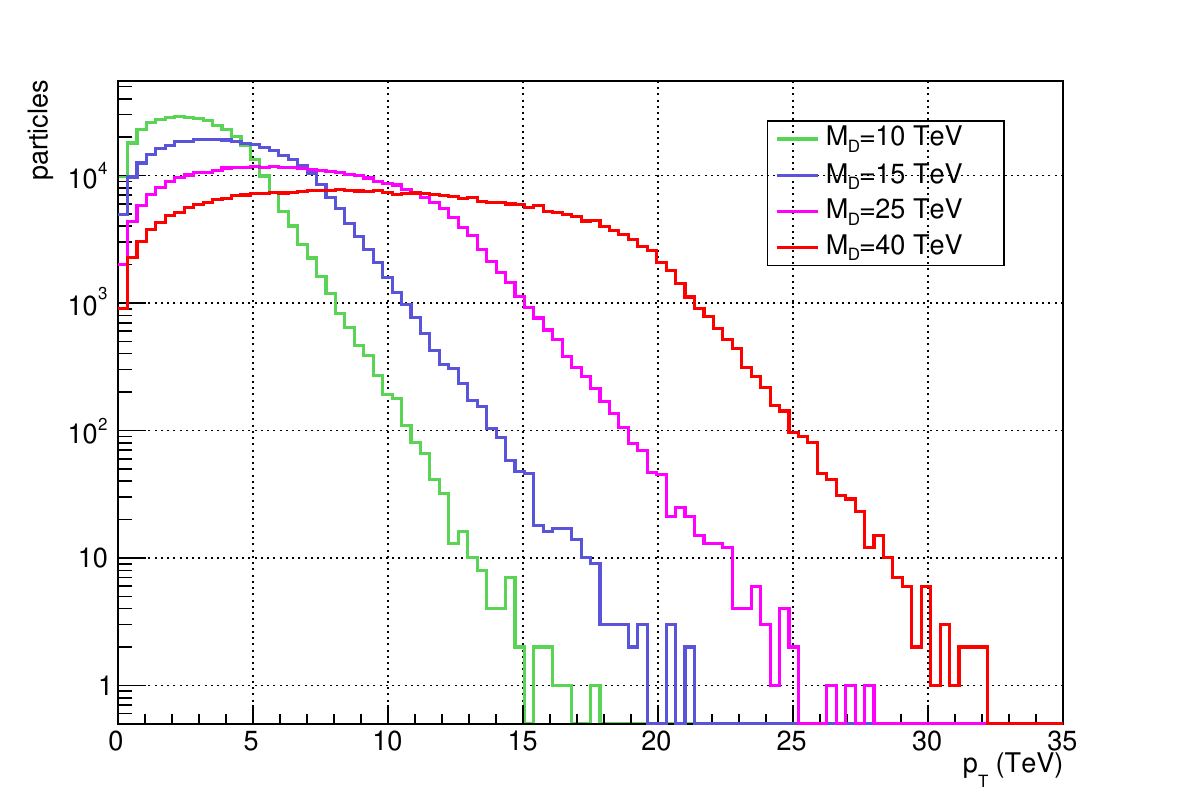}
  \caption{}
\end{subfigure}
\caption{Transverse momentum ($p_{\rm T}$) distributions of particles emitted during BH evaporation for (a) S27$\_$NR and (b) S100$\_$NR scenarios.}
\label{fig:ptcomp}
\end{figure}

\begin{figure}
\begin{subfigure}{.5\textwidth}
  \centering
  \includegraphics[width=1\linewidth]{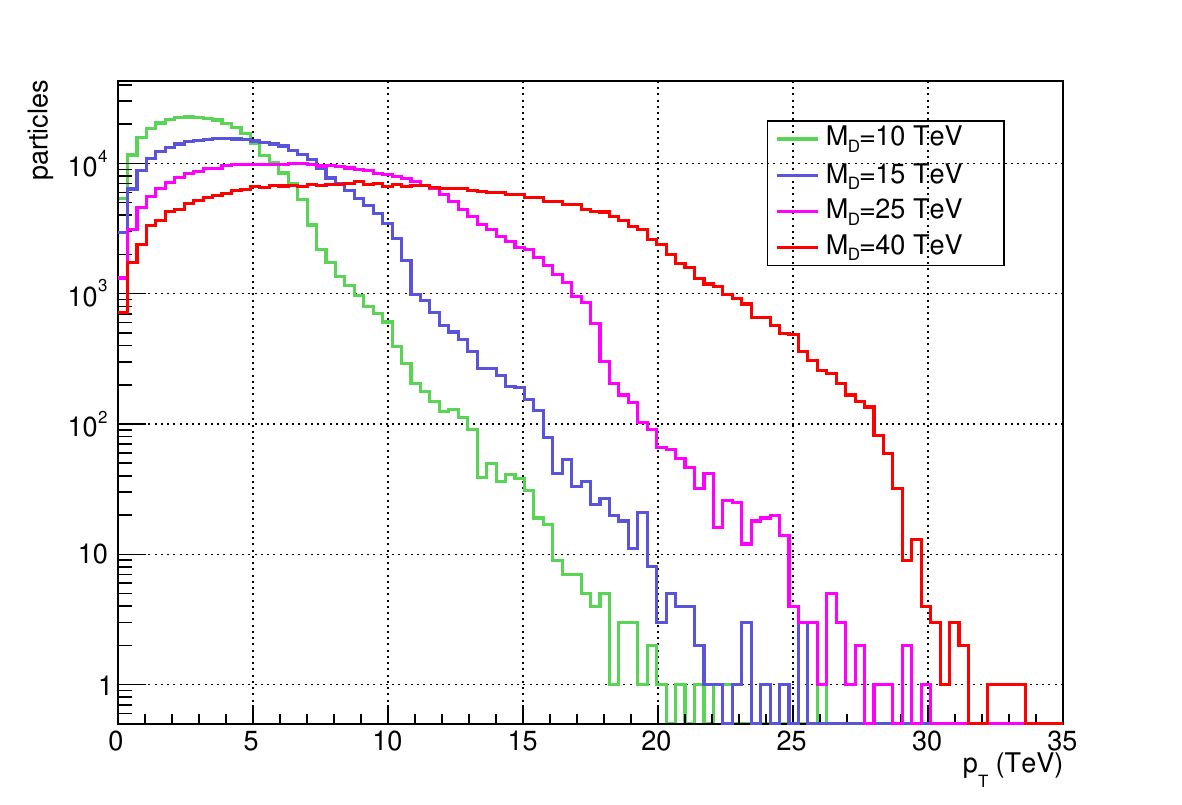}
  \caption{}
\end{subfigure}%
\begin{subfigure}{.5\textwidth}
  \centering
  \includegraphics[width=1\linewidth]{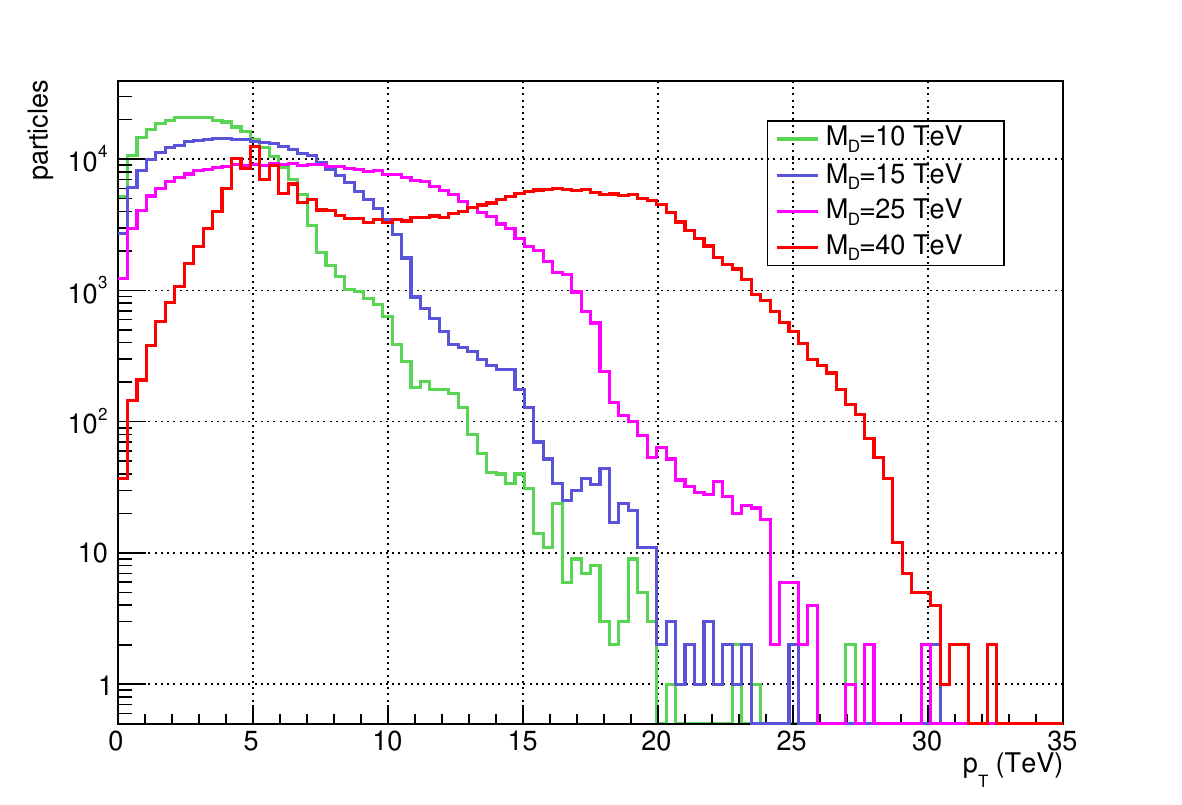}
  \caption{}
\end{subfigure}
\caption{Transverse momentum ($p_{\rm T}$) distributions of particles emitted during evaporation of rotating BHs for (a) S100$\_$R and (b) S100$\_$RL scenarios. In the S100$\_$RL case, gravitons in the final state are excluded.}
\label{fig:ptrcomp}
\end{figure}

We show transverse momentum $p_{\rm T}$ distributions of the particles radiated during BH evaporation on figure~\ref{fig:ptcomp} and figure~\ref{fig:ptrcomp} for non-rotating and rotating BH cases, respectively. $p_{\rm T}$ values go up to about 10 TeV for 27 TeV collisions, while they go up to about 30 TeV in certain 100 TeV collision energy scenarios. 


A common pattern on both collision energy cases is that particles in the final state of scenarios that belong to higher $M_{\rm D}$ tend to have higher transverse momenta. This is due to the same reason with the multiplicities' dependence on $M_{\rm D}$; squeezing of the mass distributions between $M_{\rm th}$ and available energy. Also note that distributions for the same $M_{\rm D}$ but different collision energy scenarios are different due to the difference in BH invariant mass distributions. 

\begin{figure}
\begin{subfigure}{.5\textwidth}
  \centering
  \includegraphics[width=1\linewidth]{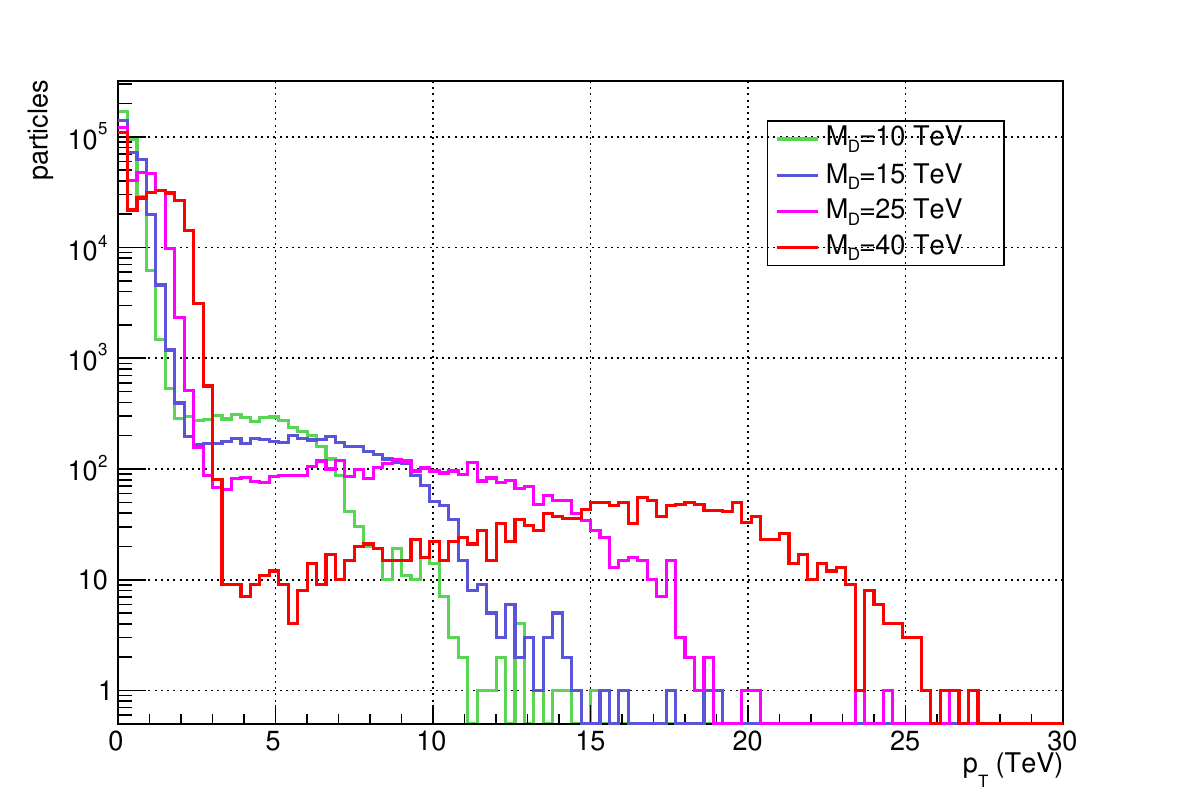}
  \caption{}
  \label{fig:grav_pt}
\end{subfigure}%
\begin{subfigure}{.5\textwidth}
  \centering
  \includegraphics[width=1\linewidth]{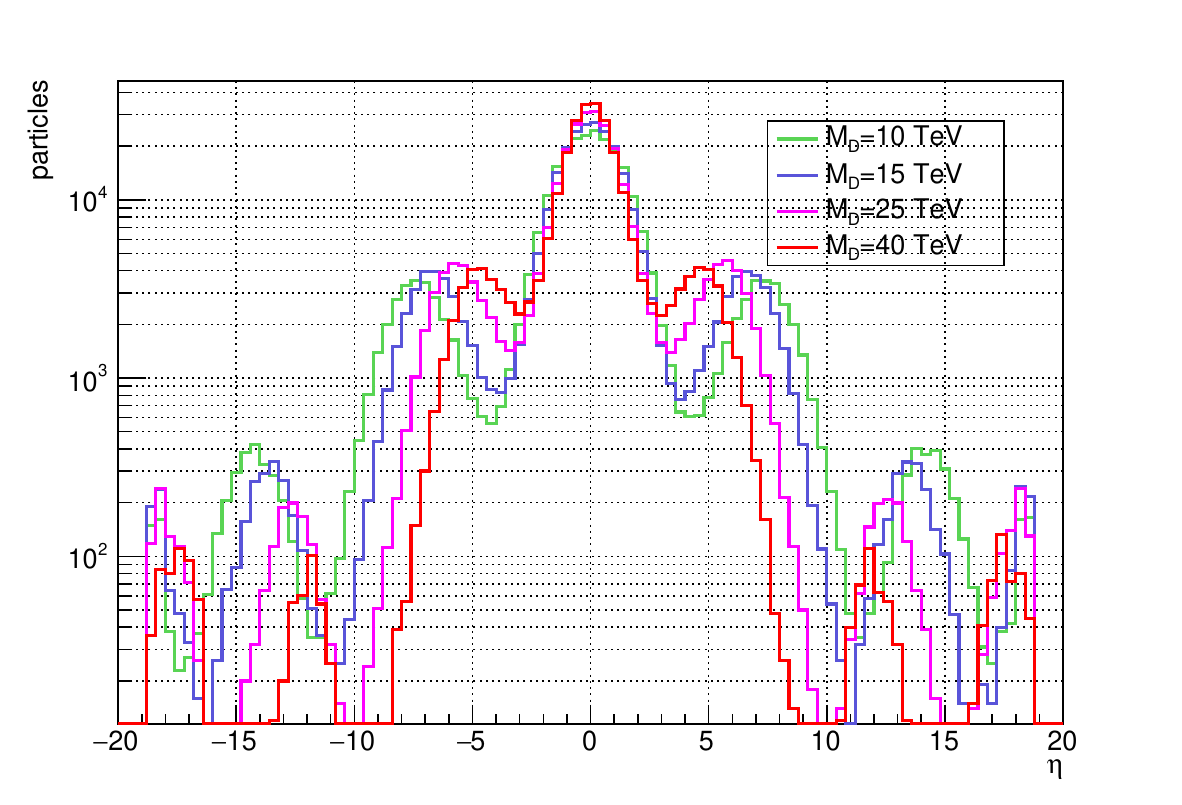}
  \caption{}
  \label{fig:grav_eta}
\end{subfigure}
\caption{$p_{\rm T}$ and $\eta$ distributions of the gravitons emitted in S100$\_$RL scenarios.}
\label{fig:grav_pt_eta}
\end{figure}

\begin{figure}
  \centering
  \includegraphics[width=0.5\linewidth]{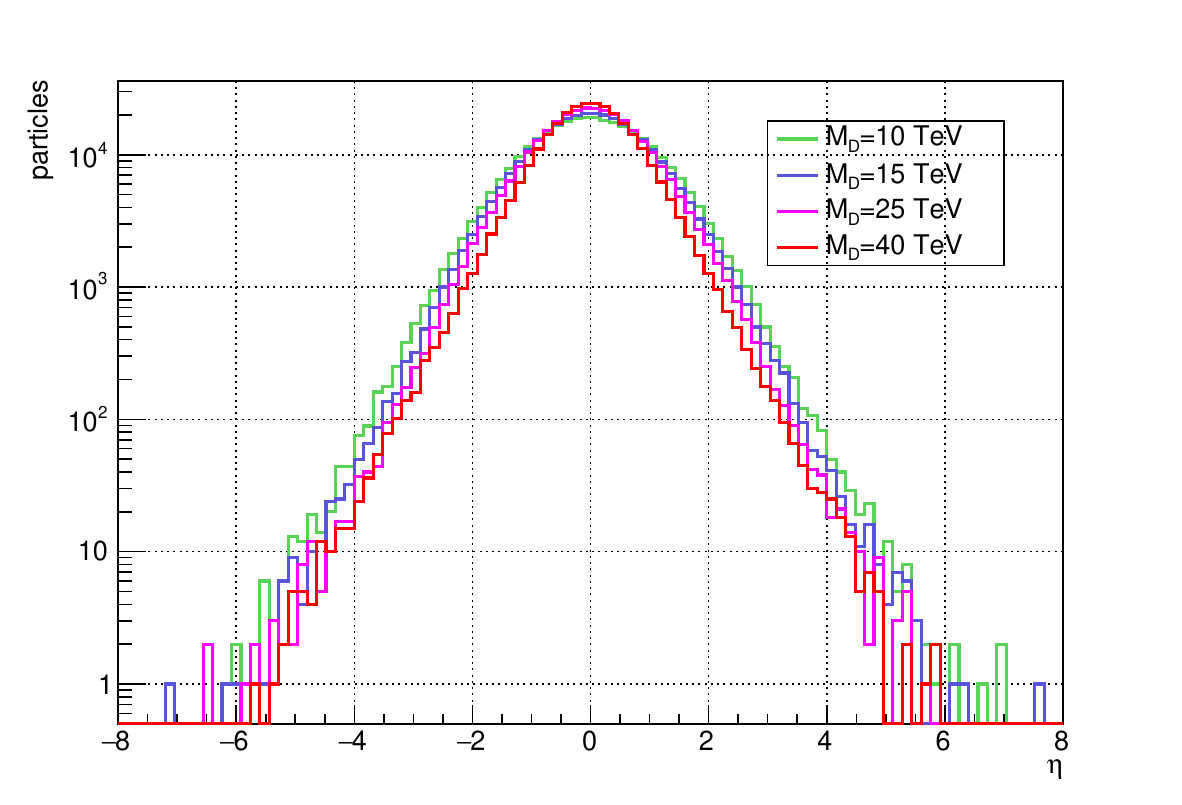}
  \caption{$\eta$ distributions for S100$\_$RL scenarios with different $M_{\rm D}$ parameter values.}
  \label{fig:comp_eta_100_r3_loss}
\end{figure}

The rotating BH distributions tell a similar story concerning transverse momenta of the emitted particles. Both the S100$\_$R and S100$\_$RL scenarios have similar PT distributions with S100$\_$NR. An unexpected dip is observed around 10 TeV in the S100$\_$RL distribution for $M_{\rm D}$=40 TeV. We histogram the kinematic distributions of the gravitons in the S100$\_$RL scenarios separately on figure~\ref{fig:grav_pt_eta}. These distributions include both energy loss gravitons (the two gravitons emitted before the BH formation), and the gravitons emitted during the BH evaporation. Here we can identify the BH evaporation gravitons from their $p_{\rm T}$ and $\eta$ values (see figure~\ref{fig:ptrcomp} and figure~\ref{fig:comp_eta_100_r3_loss} for comparison), and conclude that  the mass/momentum loss gravitons are modelled with high $\eta$ (with values up to 20) and with low $p_{\rm T}$ values by \texttt{BlackMax}. The $\eta$ distribution of gravitons is particularly interesting as it features three distinct peaks (on each side of the origin) around the central peak which are the BH evaporation gravitons.

The number of extra dimension $n$ dependence of $p_{\rm T}$ distributions were also studied and no differences to affect the general picture have been observed.

Regarding the angular distributions of emitted particles, we show pseudorapidity distributions for rotating BHs with mass/momentum loss on figure~\ref{fig:comp_eta_100_r3_loss}. We observe that almost all particles are emitted in the range $|\eta|<5$. In fact there isn't much to comment on the $\eta$ distributions of the emitted particles in various scenarios in that the distributions do not depend on collision energy, BH rotation or mass/momentum loss scenario or the number of extra dimensions $n$.

\begin{figure}
  \centering
  \includegraphics[width=0.5\linewidth]{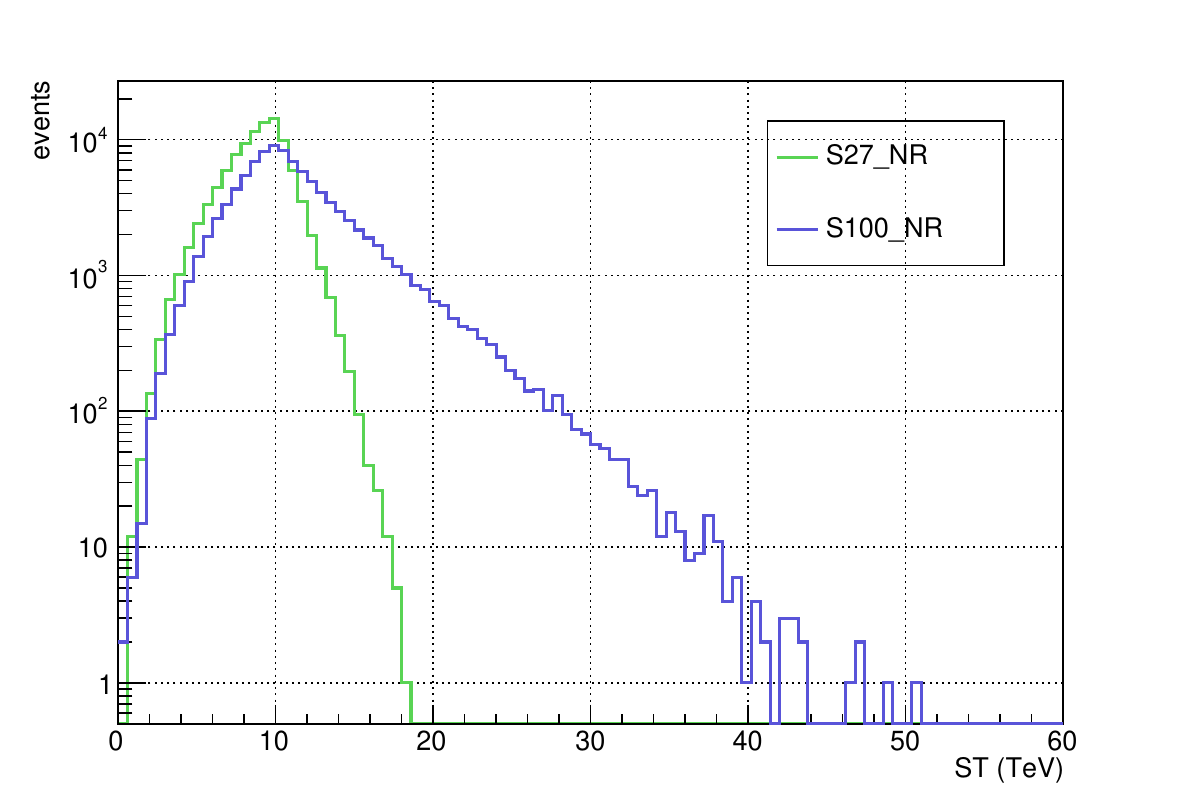}
  \caption{ST distributions for $M_{\rm D}$=10 TeV in 27 TeV and 100 TeV proton collisions for non-rotating BHs.}
  \label{fig:comp_st_r1}
\end{figure}

We can conclude this section with a few words on distinguishing microscopic BH events from possible background processes. Microscopic BHs as predicted by the ADD model seem to be on a virtually background free region, see e.g. \cite{cmsBHSph2018,atlasBH2016}. HT and ST are very effective and inclusive event variables for signal-background discrimination of microscopic BH events in LHC experiments. HT is defined as the scalar sum of the $p_{\rm T}$ of all the jets in the final state. ST includes the $p_{\rm T}$ of all final state particles, not only the jets. Missing transverse energy is also added to this variable for more realistic results. We calculated ST values for the BH evaporation final state particles\footnote{All particles including neutrinos are included in the calculation.} in the lowest $M_{\rm D}$ scenarios (figure~\ref{fig:comp_st_r1}) which show that BH events tend to have fairly high ST values compared to those of relevant SM backgrounds, e.g. QCD multijet events.
 
\section{Discussion and Conclusions}
In this article, we presented our results produced using \texttt{BlackMax} event generator for production cross-sections and final states of microscopic BH production and decay in future proton colliders with CM energies of 27 TeV and 100 TeV. Cross-section values obtained show that there is parameter space available to explore in these  colliders concerning ADD type low scale gravity models; a 27 TeV collider may explore the parameter space up to $M_D\approx15$ TeV while a 100 TeV collider may explore up to $M_D\approx45$ TeV. However, note that these values are estimates based on event generator particles and better estimates require more detailed studies that include experimental effects using particle showering and detector simulation with software tools like \texttt{Pythia} and \texttt{Delphes}. 

If formed, microscopic BHs may evaporate with a number of particles in the final state down to zero\footnote{The case with only gravitons in the final state, and no other SM particles.} and up to 17 in different parameter points. BHs tend to radiate particles in the transverse plane with high $p_T$ (with values up to 30 TeV in certain parameter points) and low pseudorapidity ($|\eta|<5$) values, which make it easier to distinguish them from background events. Together with the future colliders having access to a parameter space beyond LHC reach, it's possible for these colliders to provide much valuable information on whether large extra dimensions are part of the structure of the universe.

	\bibliography{mybibfile}
	
\end{document}